\documentclass{svjour3}                     
\smartqed  
\usepackage{graphicx}
\def\Sun{$\rm _{Sun}$}
\def\Earth{$\rm _{Earth}$}
\def\Jupiter{$\rm _{Jupiter}$}

\begin{document}

\title{Exomoon simulations}
\titlerunning{Exomoon simulations}
\author{A. E. Simon$^1$        \and
        Gy. M. Szab\'o$^1$ \and K. Szatm\'ary$^1$
}

\authorrunning{A. E. Simon,  Gy. M. Szab\'o \and K. Szatm\'ary}

\institute{   $^1$University of Szeged, Dept. Experimental Physics \&{} Astron. Obs., 6720 Szeged D\'om t\'er 9., Hungary \\
              Tel.: +36 62 544 052\\
              \email{asimon@titan.physx.u-szeged.hu}             \\
              \email{szgy@titan.physx.u-szeged.hu}\\
              \email{k.szatmary@physx.u-szeged.hu}
}

\date{Received: date / Accepted: date}
\maketitle

\begin{abstract}
We introduce and describe our newly developed code that simulates light curves and radial velocity curves for arbitrary
transiting exoplanets and its satellite. The most important feature of the program is the calculation
of radial velocity curves and the Rossiter-McLaughlin effect in such systems. We discuss the
possibilities for detecting the exomoons taking the abilities of Extremely Large Telescopes
into account. We show that satellites may be detected also by their RM effect in the future,
probably 
using less accurate measurements than promised by the current instrumental developments. Thus, RM effect will be an important observational
tool in the exploration of exomoons.

\keywords{planetary systems \and planets and satellites: general \and techniques: photometric, radial velocities \and methods: numerical}
\end{abstract}

\section{Introduction}
\label{intro}

The number of known transiting exoplanets is rapidly increasing,
which has recently inspired significant interest as to whether they can host a detectable moon
(e.g. Szab\'o et al. 2006, Simon et al. 2007, Deeg 2009).
There has been no such example where the presence of a satellite was
proven, but several methods have been investigated for such a detection
in the future (barycentric transit timing variation, Sartoretti \&{} Schneider 1999,
Kipping, 2008; photometric transit timing variation, Simon et al 2007; transit
duration variation, Kipping, 2009, Time-of-Arrival analysis of pulsars, Lewis et al. 2008).
Deviations from the perfect monoperiodic
timing of transits can be detected in $O-C$ curves (Sterken, 2005) and might suggest the presence
of a moon (D\'\i{}az et al. 2008), perturbing planets (Agol et al. 2005) or suggest periastron
precession (P\'al and Kocsis, 2008).

These methods require extremely high precision measurements, and all are based on the deep understanding
of physical effects due to a satellite. Therefore, we developed a new
code (Labview environment, NI 2008, http://www.ni.com/labview) for visualization and
precise calculations of arbitrary transiting systems with a satellite.
We simulate transit light curves and radial velocity curves for exoplanet-exomoon
systems with adjustable masses, radii, orbit periods,
eccentricities, inclinations and ascending node longitudes.
This
simulator generates light curve and radial velocity curves with Rossiter-McLaughlin
effects for arbitrary planet-satellite systems with a handful adjustable parameters.
We soon plan
to publish the program after the current developing of the peripherals
(such as improving the GU interface, enabling more input parameters,
developing fitting algorithms etc.) will have been made ready, too.

\begin{center}
\begin{table*}
\caption{Input data of the simulations}
\label{tab:1}
\begin{tabular}{lll}
\hline\noalign{\smallskip}
\hskip1cm Star mass & \hskip1cm 0.3 M\Sun \\
\hskip1cm Star radius & \hskip1cm 0.36 R\Sun \\
\hskip1cm Star rot. period &\hskip1cm 10 days \\
\hskip1cm Limb darkening &\hskip1cm 0.5 \\
\hline\noalign{\bigskip}
 \hline\noalign{\smallskip}
{\bf Simulation 3}  & {\bf Planet} & {\bf Satellite} \\
mass & 0.15 M\Jupiter & 1 M\Earth \\
radius & 0.40 R\Jupiter & 1 R\Earth \\
period &600 days &0.3 days\\
 inclination & 65$^\circ$ &80$^\circ$ \\
 asc. node & 0.04&  0$^\circ$ \\
 \hline\noalign{\smallskip}
{\bf Simulation 4}  & {\bf Planet} & {\bf Satellite} \\
 mass &0.15 M\Jupiter &1 M\Earth \\
 radius &0.45 R\Jupiter &1 R\Earth \\
 period &4300 days &0.2 days \\
 inclination &70$^\circ$ &80$^\circ$\\
 asc. node & 0.04$^\circ$& 0$^\circ$ \\
\noalign{\smallskip}\hline
\end{tabular}
\end{table*}
\end{center}

Thanks to the recent improvements 
in spectroscopic design and instrumentation 
developments, one can expect reaching the $\sim$1 
cm/s velocimetric accuracy in a foreseeable future (e.g utilizing
laser frequency combs, Li et al., 2008). Therefore
we included the calculation of $r_v$ curves into the code, and
simulated the Rossiter-McLaughlin effects of the moon. This facility is
one of the main features supported by our code, and enables to compare the
light curves and radial velocity curves simultaneously.

\section{Simulations}
\label{sec:simulations}

\begin{figure}
\hskip0.21\textwidth\includegraphics[width=0.58\textwidth]{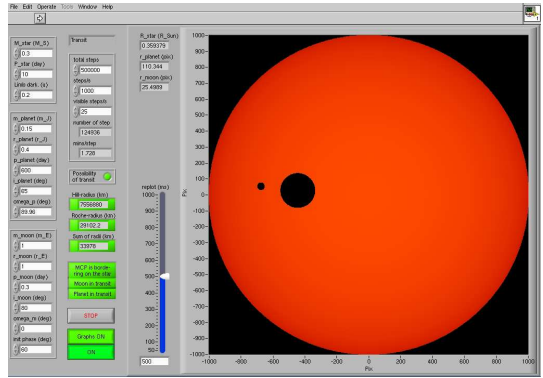}
\caption{The main GU interface of the simulator (in its present state) shows the configuration
of the transit.}
\end{figure}

\begin{figure*}
  \includegraphics[width=0.485\textwidth]{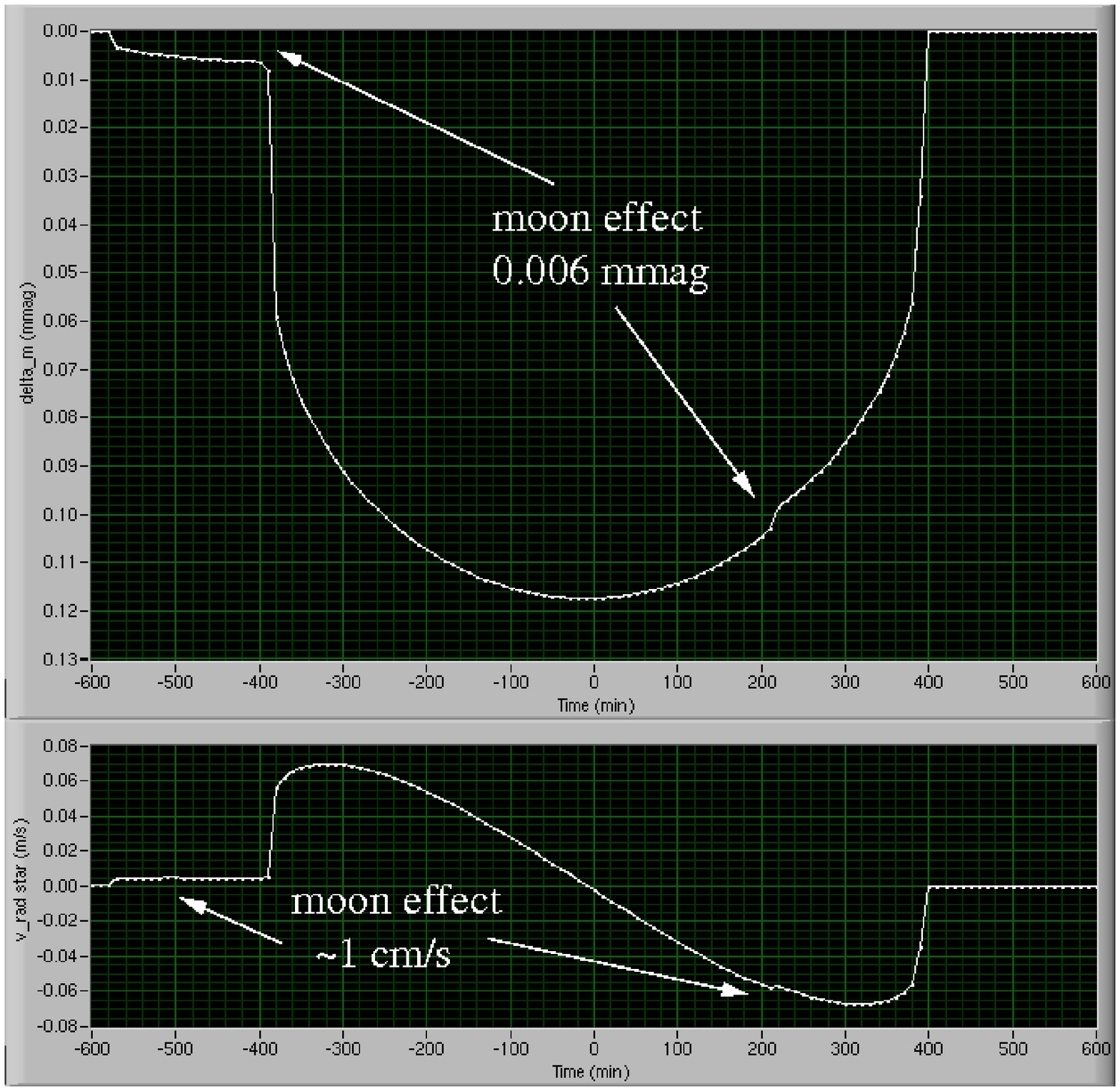}\hfill
  \includegraphics[width=0.485\textwidth]{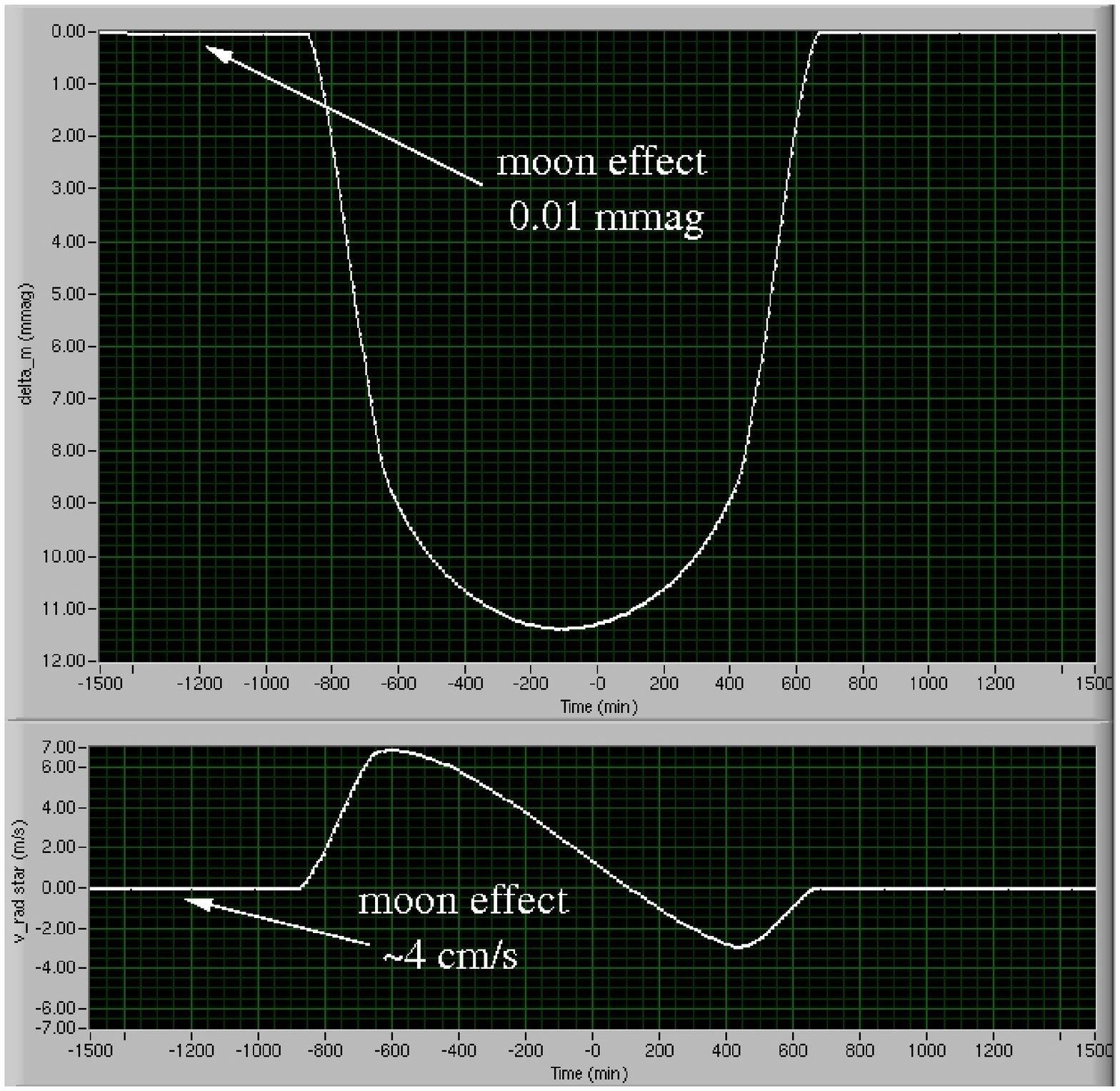}
\bigbreak
  \includegraphics[width=0.485\textwidth]{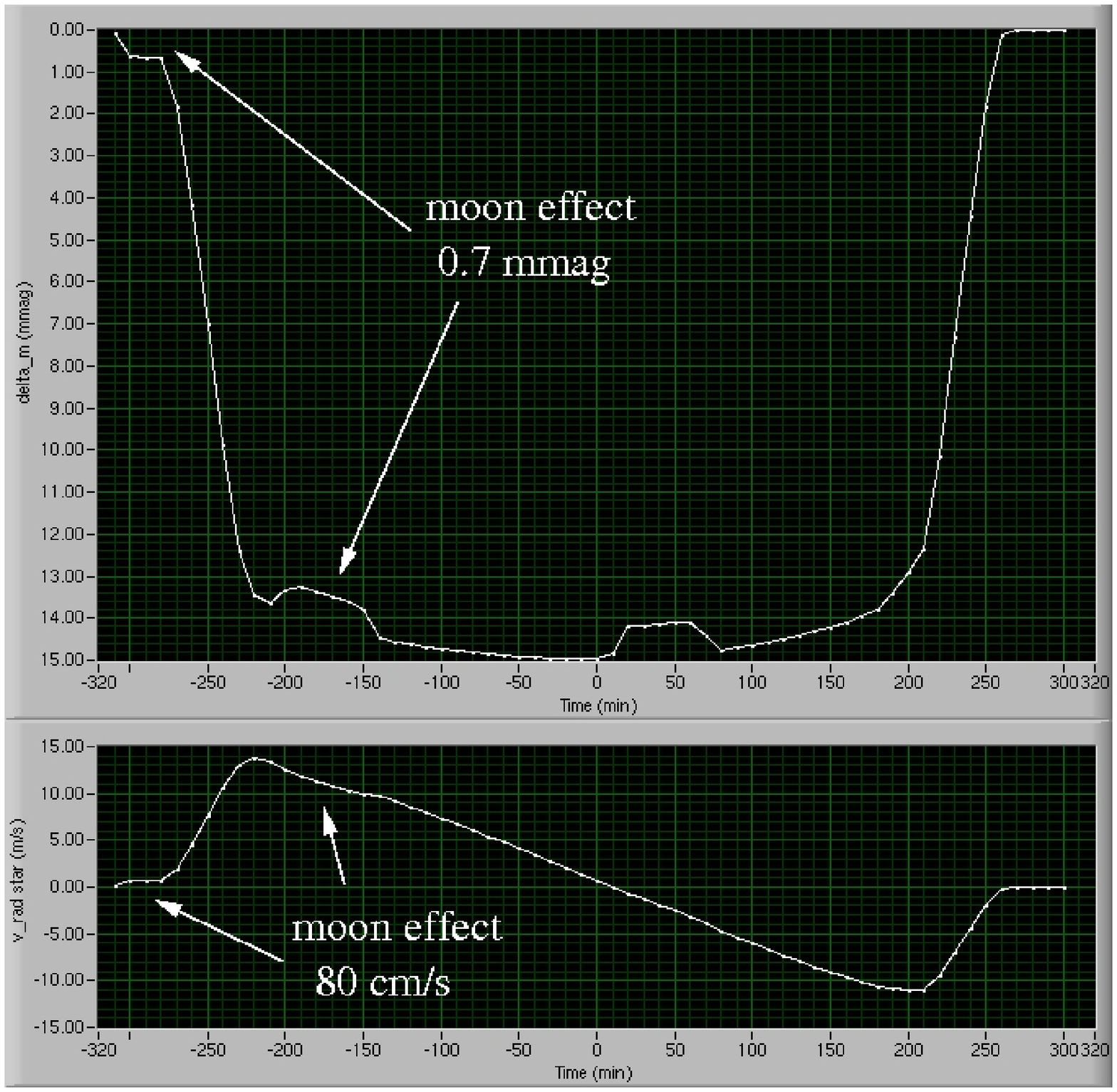}\hfill
  \includegraphics[width=0.485\textwidth]{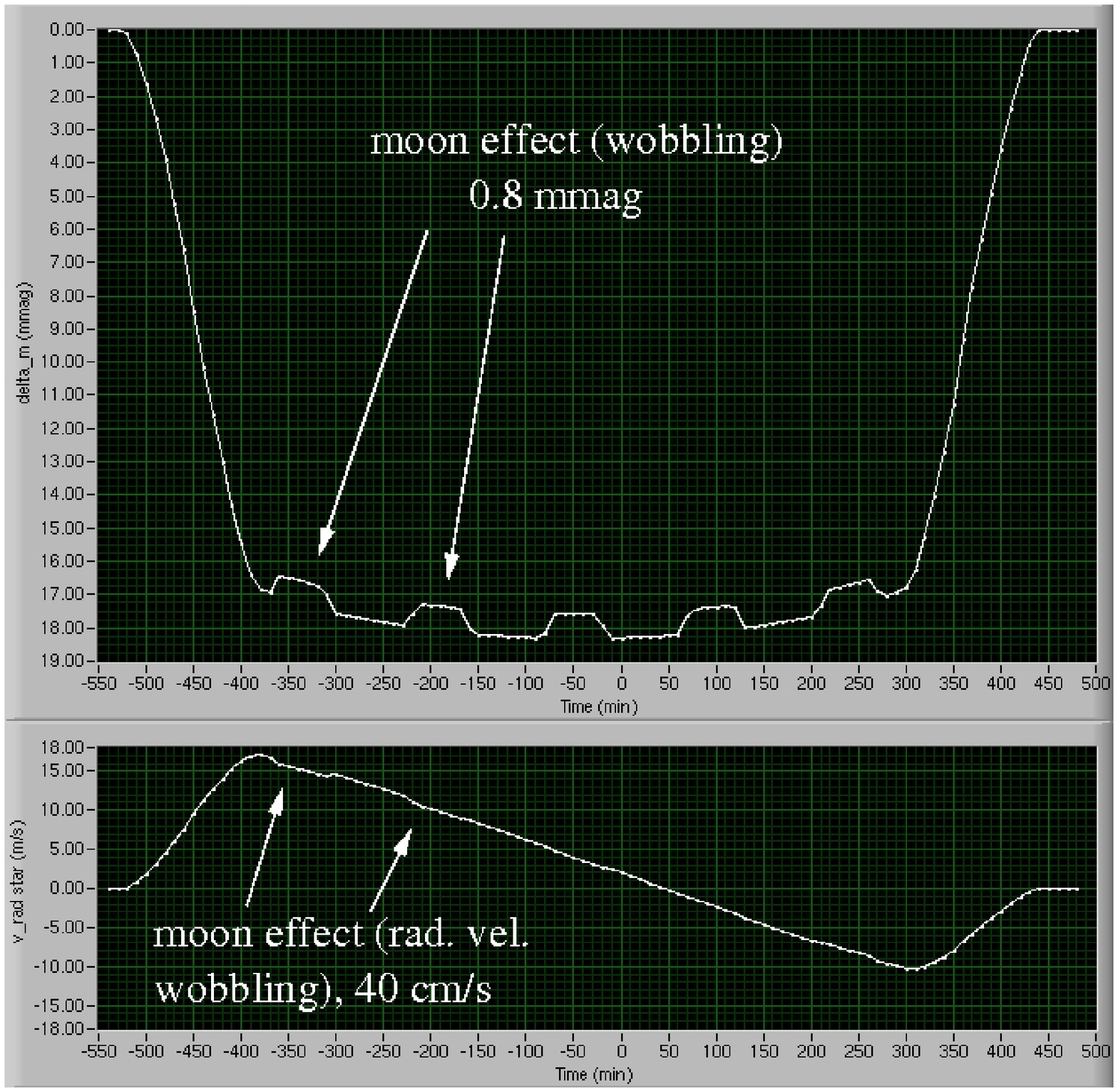}
\caption{Sample screenshots of simulations (data file GUIs). The four panels show
the light curves in the upper part, and the velocity curves below.
Top left: the Earth-Moon system; top right: the Jupiter-Ganymede system;
bottom left: Simulation 3; bottom right: Simulation 4. See Table 1
for the parameters. The simulation is for a 0.3 M$_\odot$ star.}
\label{fig:3}       
\end{figure*}

Our program offers a menu selection of four animated simulation windows (or GUIs).
These show (i) a distant view of the orbit of the planet and the satellite; (ii)
a zoom into the transit geometry (see an example in Fig 1); (iii) the surface of the star behind the transiting objects
and (iv) the light curves and radial velocity curves, respectively. In both GUIs, the transit
parameters can be adjusted. The output files contain the light curve and the radial
velocity curve with a header.

In Fig. 2 we show four sample simulations highlighting the most prominent observable light curve
and $r_v$ curve effects. The first is a modeled remote observation of our Sun-Earth-Moon
system, while the second one shows a system consisting of Jupiter and Ganymede orbiting the Sun,
also observed remotely. Two hypothetical systems were designed up with a slowly (Simulation 3) and
a rapidly (the moon orbits more times during the transit, Simulation 4) orbiting satellite.
The most prominent observable effects are marked and quantified on the screenshots. For the
hypothetical configurations we summarized the sets of system parameters in Table 1.

\section{Conclusions}
\label{sec:conclusions}

For a {\it direct detection} of the moon in the {\bf photometric light curves},
a large moon and a relatively small (thus low-mass) star is required.
There is more chance for indirect detections with
{\it Transit Timing Variation}. There is a possibility for the direct detection of a moon:
a little transient shift of the radial velocity right before (or after) the transit,
due to the {\bf Rossiter-McLaughlin effect} of the moon itself (Fig. 2 upper panels).
Note that the RM amplitudes of the Moon and
Ganymede are 0.8 cm/s and 4 cm/s, this latter is not apparent on this scaling (Sim. 2).
If the moon is so close that during one transit it orbits the planet more times, 
this can lead to marked waves (wobbling)
in the RM curve (Fig. 2 lower panels) that can have an amplitude of 10--100 cm/s.


An important observational constraint is the jitter that the
planet-hosting star displays. It will be an error source in detecting
any low-mass planets and their companions orbiting the star. Currently
there is very little available RV data with precisions better than 1 m/s,
and so there is still not sufficient data. 
The current view is that
the jitter level depends on the mass, activity and
age of the star.
Old, inactive G and K dwarfs typically
offer the best planet-detection performance: few stars
are known to have $<$1 m/s jitter, while they typically
have jitter levels in the 1--5 m/s regime (Saar 2003, Wright 2005, 
O'Toole et al. 2008). The Geneva group with HARPS
demonstrated that chromospherically quiet M dwarfs have a jitter level $\approx$ 1 m/s,
mostly due to longer time-scale phenomena such as granulation noise (Mayor et al. 2009).
Earlier-type stars generally have jitters in the 4$+$ m/s
range, while active stars often exceed 10 m/s. Younger stars have far higher jitters increasing to 700
m/s at 12 Myr age (Paulson \&{} Yelda 2006). 

The lowest jitter levels are comparable to the RM effects due to a hypothetical
satellite in the favorable cases, especially for late inactive dwarfs. Planet detection and
precise orbit modeling is also easier for these stars because of their low mass.
This means that nowadays telescopes perform nearly the precision required by a positive detections 
of exomoons, while the competent instrument for this observation will be 
the Extremely Large Telescope. 

The main conclusions can be summarized as follows.

\begin{enumerate}
\item{} Combining photometry and radial velocity measurements, there will be more
chance for detecting exomoons. The moon's effect in the RM curve was modeled for
an Earth-Moon-like and Jupiter-Ganymede-like systems. The similar extrasolar systems
may be detected in the future.
\item{} For a direct detection in photometric curves, a large moon and a relatively
small (i.e. low-mass) star is required. There is more chance for indirect
detections with Transit Timing Variation (see Simon et al. 2007) from photometry.
\item{} The radial velocity measurements can manifest the moon by its own
RM effect superimposed on the RM curve.
\item{} We suggest the
analysis of the radial velocity measurements and photometric data together
looking for signs of hosted satellites.
\end{enumerate}

\begin{acknowledgements}
This work has been supported by the OTKA K76816 Grant and the Bolyai J\'anos
Research Fellowship of the Hungarian Academy of Sciences. The traveling and living expenses 
were financed by the National Office for Research and Technology, Hungary (Mecenatura Grant), the "Lend\"ulet 
Fiatal Kutat\'oi Program" of the Hungarian Academy of Sciences and a funding provided by the Conference LOC.
\end{acknowledgements}

\end{document}